\documentclass[conference]{IEEEtran}
\IEEEoverridecommandlockouts
\usepackage{cite}
\usepackage{amsmath,amssymb,amsfonts}
\usepackage{algorithmic}
\usepackage{graphicx}
\usepackage{textcomp}
\usepackage{xcolor}
\usepackage{subcaption}
\def\BibTeX{{\rm B\kern-.05em{\sc i\kern-.025em b}\kern-.08em
    T\kern-.1667em\lower.7ex\hbox{E}\kern-.125emX}}
\begin{document}

\title{AN ANNIHILATING FILTER-BASED DOA ESTIMATION FOR UNIFORM LINEAR ARRAY
}

\author{\IEEEauthorblockN{1\textsuperscript{st} Phan Le Son}
\IEEEauthorblockA{\textit{Department of Electrical \& Computer Engineering} \\
\textit{University of Kaiserslautern}\\
Kaiserslautern, Germany \\
phan@eit.uni-kl.de}
\and
\IEEEauthorblockN{2\textsuperscript{nd} Lam Pham}
\IEEEauthorblockA{\textit{Center For Digital Safety \& Security} \\
\textit{Austrian Institute of Technology}\\
Vienna, Austria \\
lam.pham@ait.ac.at}
}

\maketitle

\begin{abstract}
In this paper, we propose a new method to design an annihilating filter (AF) for direction-of-arrival (DOA) estimation of multiple snapshots within an uniform linear array.
To evaluate the proposed method, we firstly design a DOA estimation using multiple signal classification (MUSIC) algorithm, referred to as the MUSIC baseline. 
We then compare the proposed method with the MUSIC baseline in two environmental noise conditions: Only white noise, or both white noise and diffusion. 
The experimental results highlight two main contributions; the first is to modify conventional MUSIC algorithm for adapting different noise conditions, and the second is to propose an AF-based method that shows competitive accuracy of arrival angles detected and low complexity compared with the MUSIC baseline.

\end{abstract}


\section{Introduction}

It is fact that array signal processing (ASP)~\cite{krim1996two, book_01} has been widely employed in diverse areas such as  acoustics~\cite{book_03, book_04}, radio-interferometry~\cite{book_06, book_07}, radar and sonar systems~\cite{book_01, book_02}, wireless networks~\cite{app_01, app_02, app_03} and medical imagery~\cite{book_08, book_09}.
In an ASP based system, direction-of-arrival (DOA) estimation, the process of retrieving the direction information of electromagnetic/acoustic sources by using a sensor array, is considered as one of the most important topics~\cite{krim1996two} that attracts intensive researches.
Early research on DOA mainly explored techniques of time delay estimation (TDE)~\cite{tde_01, tde_02} and steered response power (SRP)~\cite{tde_03}. 
To further improve the DOA performance, subspace-based methods such as MUSIC~\cite{schmidt1986multiple} and ESPRIT~\cite{roy1989esprit} have been widely employed.
However, the subspace-based methods  are sensitive to coherent signal~\cite{dis_01} that challenges to separate signal and noise subspaces, then leads an incorrect estimation of the spatial spectrum.
To deal with the coherent signals, various preprocessing techniques have proposed to decorrelate signals.
In particular, Pillai \textit{et al.} \cite{pillai} suggested two different spatial smoothing techniques: Forward spatial smoothing and forward backward spatial smoothing.
Recently, basing on the annihilating filter’s properties, Vetterli et al.~\cite{vetterli2002sampling} proposed the finite rate of innovation concept that reconstructs the signal perfectly from the uniform sampling.
This reconstruction concept can be directly applied to DOA estimation where the active sources act as the stream of Dirac.
However, the conventional AF-based methods are very sensitive with noise as the directions are deduced from the roots of AF after performing logarithm operations. 
Furthermore, AF-based methods aim to build a full-rank convolution matrix that requires the number of active sources less than a half of the number of measurements. 

To tackle these disadvantages of the conventional AF-based methods, we propose a design of AF-based DOA for multiple snapshots within an uniform linear array (ULA), which not only enable to detect more active sources but also is insensitive with noises.
We then compare the proposed method with the conventional MUSIC.
In addition, to consider the diffuse noise in the DOA estimation, we modify the conventional MUSIC to adapt both white noise and diffusion noise conditions.
We also examine the performance of the conventional MUSIC, modified MUSIC and proposed AF-based method under diffuse noise environments.
\section{The Extended MUSIC}
\label{sec:MUSIC}

Let us consider $N$ as the number of narrowband far-field sources, $M$ as the number of sensors, and assume both white noise and diffusion noise are uncorrelated to signal.
Then, the sound wave reaching the $m^{th}$ sensor ($0 \leq m \leq M-1$), referred to as measurement signal, is planar and modeled as
\begin{equation}
\label{e1}
r_m(\omega) = \sum_{n=0}^{N-1} a_m(\phi_n, \theta_n, \omega)s_n(\omega) + n_m(\omega)  
\end{equation}
where $\omega$ is the rotation frequency, $s_n(\omega)$ presents the strength and phase of a source signal at arrival angles $(\phi_n, \theta_n)$, $a_m(\phi_n, \theta_n, \omega)$ is the transfer function of the wave propagation from the $m^{th}$ sensor to the reference sensor, and $n_m(\omega)$ is the additive noise. 
Notably, the $\omega$ is omitted in the remaining of this paper for conciseness.

Given the sensor array, (\ref{e1}) can be presented by a measurement vector as
\begin{equation}
\label{e2}
\begin{bmatrix}r_0 \\ r_1 \\ \cdots \\ r_{M-1}\end{bmatrix} = \begin{bmatrix} \mathbf{a}_0 & \mathbf{a}_1 & \cdots & \mathbf{a}_{N-1} \end{bmatrix}\begin{bmatrix}s_0 \\ s_1 \\ \cdots \\ s_{N-1}\end{bmatrix} + \begin{bmatrix}n_0 \\ n_1 \\ \cdots \\ n_{M-1}\end{bmatrix}
\end{equation}
where $\mathbf{a}_n$ $(0 \leq n \leq N-1)$ is the steering vector at the $n^{th}$ direction. Then, (\ref{e2}) can be presented in a sort form as
\begin{equation}
\label{e3}
\mathbf{r} = \mathbf{A}\mathbf{s} + \mathbf{n}.
\end{equation}
Given the vector $\mathbf{r}$, covariance matrix $\mathbf{R}$ is defined as
\begin{equation}
\label{e4}
\mathbf{R} = E[\mathbf{r}\mathbf{r}^H]
\end{equation}
where $E[.]$ is the expectation operation. Then, (\ref{e3}) can be substituted by
\begin{equation}
\label{e5}
\mathbf{R} = E[(\mathbf{A}s + \mathbf{n}) (\mathbf{A}s + \mathbf{n})^H ].
\end{equation}
Suppose that signal and noise are uncorrelated, (\ref{e5}) then becomes
\begin{equation}
\label{e6}
\mathbf{R} = \mathbf{A}\mathbf{S}\mathbf{A}^H + \mathbf{N}
\end{equation}
where $\mathbf{N} = E[\mathbf{nn}^H]$ is noise covariance matrix and  $\mathbf{S} = E[\mathbf{ss}^H]$ is the covariance matrix of source signals.

As we assume noise signal includes white noise and diffuse noise, $\mathbf{N}$ can be presented as
\begin{equation}
\label{e7}
\mathbf{N} = \sigma^2_d \mathbf{\Gamma} + \sigma^2_w \mathbf{I}
\end{equation}
where $\mathbf{I}$ is the identity matrix of white noise, $\mathbf{\Gamma}$ is the correlation matrix of diffuse noise, $\sigma^2_d$ and $\sigma^2_w $ are the power of diffuse noise and white noise, respectively. 
Notably, the conventional MUSIC algorithm dose not consider the diffusion noise that may negatively affect the performance. 

Given both noise conditions, (\ref{e6}) can be presented as
\begin{equation}
\label{e8}
\mathbf{R} = \mathbf{A}\mathbf{S}\mathbf{A}^H +  \sigma^2 \mathbf{N}_{v}
\end{equation}
where $\sigma^2 = \sigma^2_d + \sigma^2_w$, $\mathbf{N}_{v} = \frac{\sigma^2_d \mathbf{\Gamma}  + \sigma^2_w \mathbf{I}}{\sigma^2_d + \sigma^2_w} =  {\alpha \mathbf{\Gamma} + \mathbf{I}\over \alpha + 1}$ and $\alpha = {\sigma^2_d / \sigma^2_w}$ is a ratio representing for the noise model.
As $\mathbf{N}_v$ is a symmetric matrix, (\ref{e8}) then becomes
\begin{equation}
\label{e9}
\mathbf{R'} := \mathbf{R}\mathbf{N}_{v}^{-1} = \mathbf{A}\mathbf{S}\mathbf{A}^H \mathbf{N}_{v}^{-1} + \sigma^2 \mathbf{I}.
\end{equation}

Now we decompose $\mathbf{R'}$ to signal subspace and noise subspace by finding the eigenvalues and eigenvectors of $\mathbf{R'}$  basing on the amplitude of eigenvalues as presented by (\ref{e9_1}). 
\begin{equation}
\label{e9_1}
\mathbf{R'} = \mathbf{V}\mathbf{D}\mathbf{V}^{-1}
\end{equation}
where $\mathbf{V}$ is formed by eigenvectors of $\mathbf{R'}$ and $\mathbf{D}$ is diagonal matrix of eigenvalues. $\mathbf{D}$ can be presented as 
$$
\mathbf{D} = \begin{bmatrix} 
\lambda_1 + \sigma^2_1 & 0           & \cdots                 & \cdots         & \cdots      & 0\\ 
0                      & \cdots      &\cdots                  & \cdots         & \cdots      & \cdots\\
\cdots                 & \cdots      & \lambda_N + \sigma^2_N & \cdots         & \cdots      & \cdots\\
\cdots                 & \cdots      & \cdots                 &\sigma^2_{N+1}  & \cdots      & \cdots\\
\cdots                 & \cdots      &\cdots                  &\cdots          & \cdots      & \cdots\\
0                      & \cdots      &\cdots                  &\cdots          & \cdots          & \sigma^2_M\\
\end{bmatrix}
$$
where $\lambda_1,\dots,\lambda_N,0,\dots,0$ are the eigenvalues describing the signal subspace,  $\sigma_1^2, \dots, \sigma^2_M$ are the eigenvalues of the noise subspace.
As $\mathbf{A}\mathbf{S}\mathbf{A}^H \mathbf{N}_{v}^{-1}$ is a semi-positive definite matrix, its eigenvalues are non-negative ($\lambda_1,\dots,\lambda_N > 0$). 
In theory, $\sigma_1^2 = \dots = \sigma^2_M =  \sigma^2_w$,  but they are a set of small values in practices. 
Based on the amplitude of eigenvalues, we can separate the eigenvector of $\mathbf{R'}$ into noise subspace $\mathbf{V}_N$ and signal subspace $\mathbf{V}_S$ denoted as
$$
\mathbf{V} = [\mathbf{V}_S, \mathbf{V}_N].
$$

Note that $\mathbf{V}$ is an unitary matrix, then noise subspace is orthogonal to signal subspace. We can also explain this property by simple modified in the equations, that is, taking an column vector $\mathbf{v}_i$ in noise subspace $\mathbf{V}_N$ and multiplying it to both sides of (\ref{e6}) we obtain
$$
\mathbf{R'}\mathbf{v}_i = (\mathbf{A}\mathbf{S}\mathbf{A}^H \mathbf{N}_{v}^{-1} +  \sigma^2 \mathbf{I})\mathbf{v}_i $$
$$
{\sigma^2} \mathbf{v}_i = \mathbf{A}\mathbf{S}\mathbf{A}^H \mathbf{N}_{v}^{-1}\mathbf{v}_i + {\sigma^2}\mathbf{v}_i
$$ 
$$
\mathbf{A}\mathbf{S}\mathbf{A}^H \mathbf{N}_{v}^{-1}\mathbf{v}_i = 0
$$
\begin{equation}
\label{e10}
\mathbf{A}^H (\mathbf{N}_{v}^{-1}\mathbf{v}_i) = 0.
\end{equation}
Let us define $\mathbf{v'}_i := \mathbf{N}_{v}^{-1}\mathbf{v}_i$, (\ref{e10}) is then presented as
$$
\mathbf{A}^H \mathbf{v'}_i = 0,
$$
then
\begin{equation}
\label{e11}
\mathbf{a}^H_n\mathbf{v'}_i = 0.
\end{equation}
As (\ref{e11}) is true for all column vectors of the noise subspace $\mathbf{V}_N$, we have $\mathbf{a}^H_n\mathbf{V'}_N = 0$ where $\mathbf{V'}_N= \mathbf{N}_{v}^{-1}\mathbf{V}_N$.
Based on conventional MUSIC algorithm, we then suggest main steps below to design an extended MUSIC algorithm that takes the diffuse noise into account, that is, the noise model ratio $\alpha = {\sigma^2_d / \sigma^2_w}$ is considered as an input of the new algorithm. 
\begin{itemize}
	\item We firstly perform the decomposition of eigenvalues on $\mathbf{R'} = \mathbf{R}( {\alpha \mathbf{\Gamma} + \mathbf{I}\over \alpha + 1})^{-1}$ to obtain the non-increasing eigenvalue $\lambda_1 + \sigma^2_1  \geq \dots \geq \lambda_N+ \sigma^2_N \geq \dots \geq \lambda_M$. 
	\item Based on the amplitude of eigenvalues, we separate the corresponding eigenvectors into two groups:  The first group of signal subspace $\mathbf{V}_S = [\mathbf{v}_{1}, \dots \mathbf{v}_N]$ and the second group of noise subspace $\mathbf{V}_N = [\mathbf{v}_{N+1}, \dots \mathbf{v}_M]$. 
	\item We then modify the noise subspace to consider the diffuse noise 
	$$\mathbf{V'}_N= \big( {\alpha \mathbf{\Gamma} + \mathbf{I}\over \alpha + 1}\big)^{-1} \mathbf{V}_N.$$
	
	\item Thus, we construct the power spectrum function as
	$$
	P(\mathbf{a}_i) = \frac{1}{\mathbf{a}^H_i \mathbf{V'}^H_N \mathbf{V'}_N \mathbf{a}_i}.
	$$
	\item Finally, we search the peaks of $P(\mathbf{a}_i)$ to detect the active sources.
\end{itemize}

\section{Annihilating Filter-Based Method For DOA}
\label{sec:AF}

For ULA, each steering vector can be presented as
\begin{equation}
\label{e21}
 \mathbf{a}_n = [a^0_n, a^1_n,\dots,a^{M-1}_n]^T ~~~~  0  \leq n \leq N-1
\end{equation}
where $a_n = e^{-j\omega d_H \cos\theta_n/c}$, $d_H$ is inter-distance between two sensors, and $c$ is wave speed. 
If  $N \neq 0$, then the sound wave reaching the array measured by $\mathbf{r}$ in (\ref{e1}) is the linear combination of $N$ complex exponential vectors $\mathbf{a}_n$.
Let us define a filter with $z$-transform as
$$
F(z) = \sum_{n=0}^{N} F[n]z^{-n},
$$
which has $N$ zeros at $a_n =  e^{-j\omega d_H \cos\theta_n/c}, \forall n = 0,\dots, N-1$. Then, $F(z)$ can be presented by
$$
F(z) = \prod_{n=0}^{N-1}(1-a_n z^{-1})
$$
Note that $F[n] \; ( 0 \leq n \leq N)$ is the convolution of $N$ first-order filters with coefficients $[1, - a_n]$. 
It is easy to observe that $[1, - a_n]*\mathbf{a}^T_n = \mathbf{0}$. 
Therefore, the defined filter $F(z)$ suppresses the directional signals in the measurement signal, which reasons why the filter is called Annihilating Filter (AF) \cite{vetterli2002sampling}.
Applying the AF to the measurement signals $\mathbf{r}$, we have 
\begin{equation*}
\begin{aligned} &
\big[ F[0],\dots,F[N] \big]*\mathbf{r}^T  \\&
= [1, - a_1]*\dots*[1, - a_N] *(\sum_{n=0}^{N-1}\mathbf{a}^T_n s_n + \mathbf{n}^T)   \\& 
= (\sum_{n=1}^{N} s_n [1, - a_1]*\dots*[1, - a_N]*\mathbf{a}^T_n ) \\& 
\qquad \qquad + [1, - a_1]*\dots*[1, - a_N]*\mathbf{n}^T  \\& 
=  \mathbf{0} + [1, - a_1]*\dots*[1, - a_N]*\mathbf{n}^T \\& 
=\big[ F[0],\dots,F[N] \big]*\mathbf{n}^T.
\end{aligned}
\end{equation*}

Given the definition of $F(z)$, we know $F[0]=1$ and $M \geq N + 1$ to complete the convolution. 
In the case of noiseless ($\mathbf{n} = 0$), we have 
\begin{equation}
\label{e22}
\big[ F[0],\dots,F[N] \big]*\mathbf{r} = 0.
\end{equation}

Given the measurement signal of the array, finding the coefficients of the filter $F[n]$ can be solved by constructing the equations as shown in (\ref{equ:AF_equs}), which are deduced from (\ref{e22}). 
\begin{equation}
\label{equ:AF_equs}
\begin{bmatrix}
r_0 & r_1 & \cdots & r_{N} \\
r_1 & r_2 & \cdots & r_{N+1} \\
\cdots & \cdots & \cdots & \cdots \\
r_{N-1} & r_{N} & \cdots & r_{2N-1} \\
\end{bmatrix}
\begin{bmatrix}F[N] \\ \cdots \\ F[1] \\ F[0]\end{bmatrix} 
=   \begin{bmatrix}0 \\ 0 \\ \cdots \\ 0\end{bmatrix}.
\end{equation}

If we assign $F[0]=1$, then (\ref{equ:AF_equs}) becomes
\begin{equation}
\label{equ:AF_equs1}
\begin{bmatrix}
r_0 & r_1 &\cdots & r_{N-1} \\
r_1 & r_2 &\cdots & r_{N} \\
\cdots & \cdots & \cdots & \cdots \\
r_{N-1} & r_{N} & \cdots & r_{2N-2} \\
\end{bmatrix}
\begin{bmatrix}F[N] \\ \cdots \\ F[2] \\ F[1]\end{bmatrix} 
=   -\begin{bmatrix}r_{N} \\ r_{N+1} \\ \cdots \\ r_{2N-1}\end{bmatrix}.
\end{equation}

Equation (\ref{equ:AF_equs1}) has a unique solution mentioned in~\cite{van1983matrix}, then set of $F[n]$ is unique. 
After solving (\ref{equ:AF_equs1}), we find the roots of $F(z) = \sum_{n=0}^{N} F[n]z^{-n}$, then obtain $a_0,\dots,a_{N-1}$.
Finally, the direction of active sources can be achieved by
\begin{equation}
\label{equ:sol_DOA}
\theta_n =  \arccos {j c \log a_n \over \omega d_H}, \forall n= 0,\dots, N-1. 
\end{equation}
In order to achieve (\ref{equ:AF_equs1}), there are two considered constrains.
Firstly, the number of sensors is greater than or equal to two times the number of sources ($M \geq 2N$).
Secondly, SRN needs to be very high to assure $\mathbf{n} \approx 0$. 

Furthermore, the roots of AF associated with the true DOAs stay on the unit circle.
Therefore, we can utilize this property to evaluate the $a_n$ as  
\begin{equation}
\label{equ:check_DOA}
\vert \text{Re}\{\log a_n\} \vert \leq \beta
\end{equation} 
where $\text{Re}\{.\}$ is the real component of a complex number, $\vert . \vert $ is the absolute operator and  $\beta$ is a small value (e.g. $\beta = 0.02$).  The inequality (\ref{equ:check_DOA}) is used to select the reliable $a_n$, thus we can estimate the DOA without knowing the number of DOAs in advance. To deal with different SNR levels, we could decrease or increase $\beta$ to compromise between the accuracy and the robustness of the algorithm. 

In summary, the method in \cite{vetterli2002sampling} and the constraint in (\ref{equ:check_DOA}) can apply for the DOA estimation of coherent signals.
However, the number of sources is limited and the result is sensitive to the noise.   
In order to detect more DOAs in the noise environment, we apply a similar idea of the AF design, but for multiple snapshots. 
Suppose that the signal of active sources are frame-variant, that means the strength and phase of the signals are then varied over frame.
It leads that the signals at different snapshots are almost independent. 
This assumption is reasonable for many applications (e.g. audio, radar, etc.).
Note that the incoherent signals need to be frame-variant. Therefore, the assumption of frame-variance is automatically true for incoherent signals.
Similarly to (\ref{equ:AF_equs1}), let us build the equations for the AF from $K$ snapshots as:

\begin{equation}
\label{equ:AF_mulFrames}
\begin{bmatrix}
\mathbf{r'}^T_1 \\ 
\mathbf{r'}^T_2 \\
\dots \\
\mathbf{r'}^T_K \\
\end{bmatrix}
\begin{bmatrix}F[M-1] \\ \cdots \\ F[2] \\ F[1]\end{bmatrix} 
=   -\begin{bmatrix}r_{M,1} \\ r_{M,2} \\ \cdots \\ r_{M,K}\end{bmatrix}
\end{equation}
where $\mathbf{r'}_k, (\forall k=1,\dots,K$) is the measurement signals at snapshot $k$ after removing the last value $r_{M,K}$ (e.g. the value of the last sensor). 
Then, we can solve $F[n]$ from least-mean-square error sense as
\begin{equation}
\label{equ:AF_mulFrames_sol}
\begin{bmatrix}F[M-1] \\ \cdots \\ F[2] \\ F[1]\end{bmatrix} 
=   -(
\mathbf{X'}^H \mathbf{X'})^{-1} \mathbf{X'}^H 
\begin{bmatrix}r_{M,1} \\ r_{M,2} \\ \cdots \\ r_{M,K}\end{bmatrix}
\end{equation}
where $\mathbf{X'}= [ \mathbf{r'}_1, \mathbf{r'}_2, \dots \mathbf{r'}_K]^T$.
The solution in (\ref{equ:AF_mulFrames_sol}) is robust against noise and it is possible to detect maximum $M-1$ sources. 
In practice, $(\mathbf{X'}^H \mathbf{X'})^{-1}$ can be updated iteratively over the frame to reduce the complexity of the inverse operation. By applying Woodbury formula \cite{woodbury1950inverting}, we have
$$
(\mathbf{X'}^H_{k+1} \mathbf{X'}_{k+1})^{-1} = (\mathbf{X'}^H_{k} \mathbf{X'}_{k}+\mathbf{r'}_{k+1}\mathbf{r'}^H_{k+1})^{-1} 
$$
$$
= \mathbf{B}_k^{-1} - \mathbf{B}_k^{-1}\mathbf{r'}_{k+1}
(\mathbf{I}+\mathbf{r'}^H_{k+1}\mathbf{B}_k^{-1}\mathbf{r'}_{k+1})
\mathbf{r'}^H_{k} \mathbf{B}_k^{-1}  
$$
where $\mathbf{B}_k = \mathbf{X'}^H_{k} \mathbf{X'}_{k}$ is the matrix of $\mathbf{X'}^H \mathbf{X'}$ at the frame $k$. 
The computation of $\mathbf{B}_{k+1}^{-1}$ has complexity $\mathcal{O}(M^2)$, then (\ref{equ:AF_mulFrames_sol}) has total complexity $\mathcal{O}(KM^2)$.
After obtaining the AF coefficients $F[n]$, applying the similar approach to the conventional AF-based technique (\ref{equ:sol_DOA}) and (\ref{equ:check_DOA}) to estimate the DOA.

\section{Numerical Simulations}
\label{sec:sim}
To evaluate the proposed AF-based method, we separated our simulations into two main parts basing on the noise conditions: Simulations with only white noise, and simulations with both white noise \& diffusion noise.
In all simulations, the number of multiple snapshot is constant set to $K = 100$. 
The number of sensors is also constant set to $M =  11$ with the constrain of half-wavelength inter-distance of sensors.
Regarding the metric for evaluating, we use the  benchmark root-mean-squared error (RMSE) criteria defined as
\begin{equation}
\label{metric}
E = \sqrt{{1 \over N}\sum_{i=1}^{N}(\phi_i -\bar{\phi}_i)^2}
\end{equation}
where $N$ is the number of sources, $\bar{\phi}_i$ and ${\phi}_i$ denote estimated DOAs and the true DOAs, respectively.

\subsection{Simulations With Only White Noise}
\begin{figure}[t]
	\centering 	
	\begin{subfigure}{0.9\linewidth}
		\includegraphics[width=\linewidth]{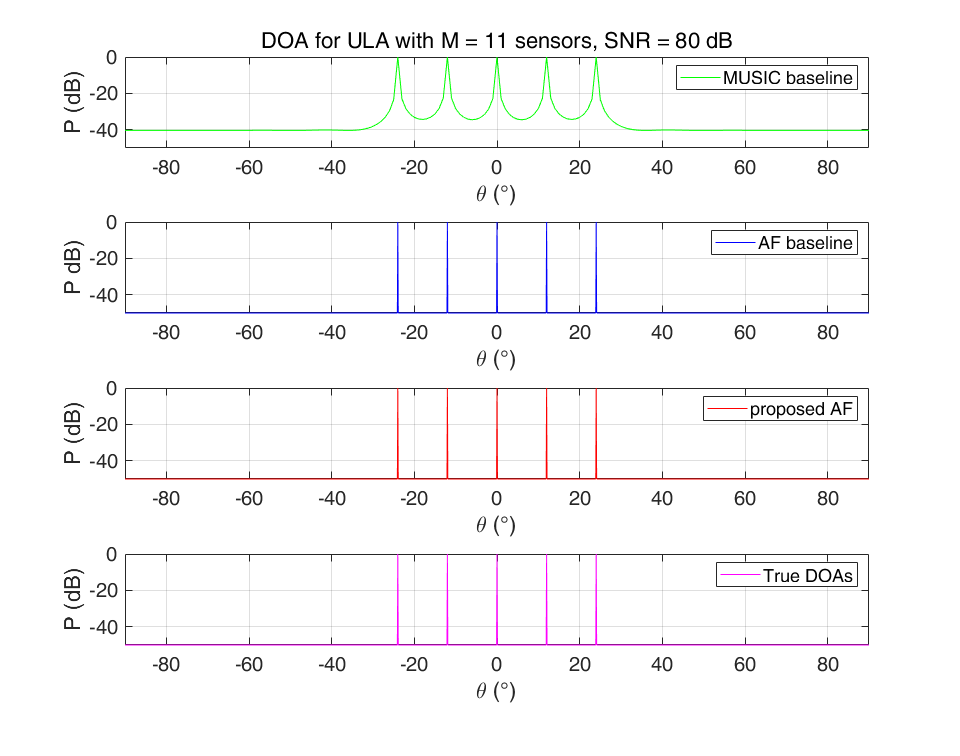}
	     \vspace{-0.8cm}
      	\caption{SNR = 80 dB.}
		\label{fig:1a}
	\end{subfigure}	
	\begin{subfigure}{0.9\linewidth}
		\includegraphics[width=\linewidth]{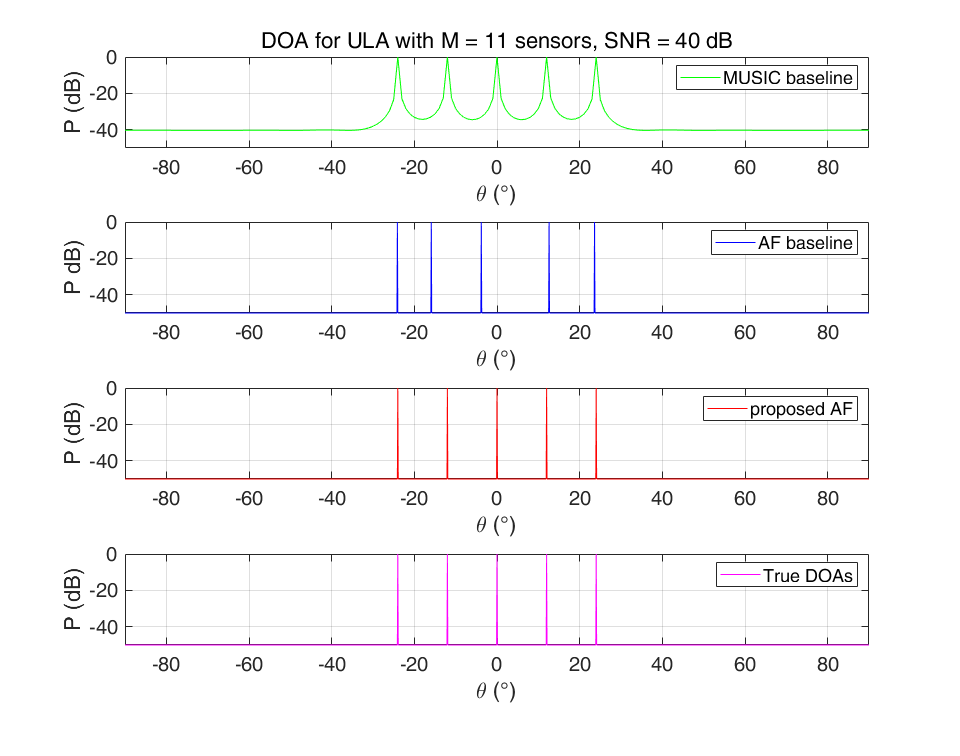}
		     \vspace{-0.8 cm}
		\caption{SNR = 40 dB.}
		\label{fig:1b}		
	\end{subfigure}
	\caption{Power spectrum comparison with N=5, white noise, and SNR reduces from 80 dB to 40 dB. }
	\label{fig:snr}
\end{figure} 
\begin{figure}[t]
	\begin{center}
		\includegraphics[width=8cm]{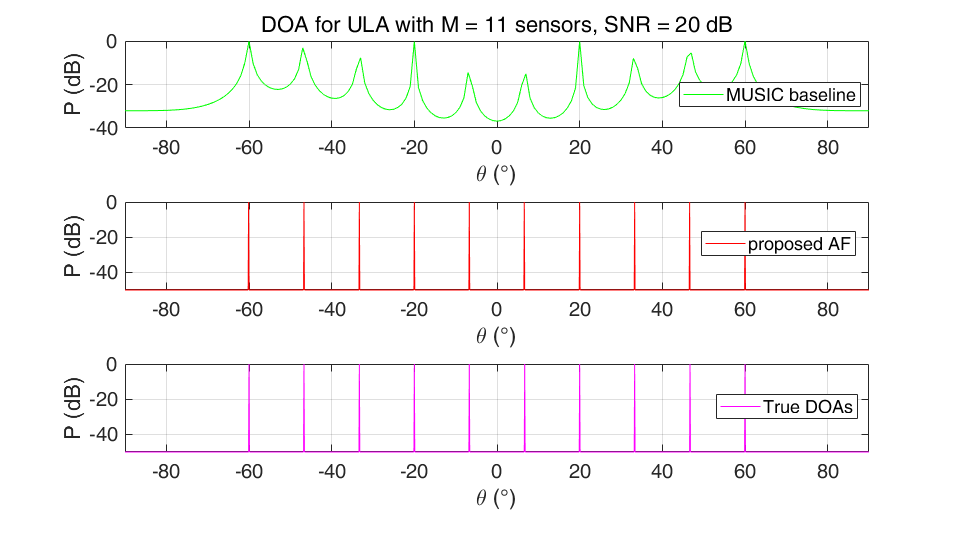}
	\end{center}
    \vspace{-0.6 cm}
	\caption{Power spectrum comparison with white noise,  N = 10.}
     \vspace{-0.4 cm}
	\label{fig:f2}
\end{figure}
Given the assumption of only white noise, we firstly evaluate how SNR affects the AF -based method (AF baseline) 's performance.
Note that this assumption makes (\ref{e7}) become
\begin{equation}
\label{e7_1}
\mathbf{N} = \sigma^2_w \mathbf{I}
\end{equation}
To this end, we conduct an experiment with the setting: The number of incoherent sources is set to $N=5$ with incident angles of $\phi_0 = -24^{o}$, $\phi_1 = -12^{o}$, $\phi_2 = 0^{o}$, $\phi_3 = 12^{o}$ and $\phi_4 = 24^{o}$, and the SNR is set to 80 dB or 40 dB.
As the results are shown in Fig.~\ref{fig:snr}, when the SNR drops from 80 dB to 40 dB, the RMSE of AF-based method with single snapshot increases from $\approx 0^o$ to $2.5^o$.
However, the MUSIC baseline and the AF-based method with multiple snapshots (proposed AF) show competitive, achieve the RMSE scores of $\approx0^o$, regardless the reduce of SNR. 

To evaluate whether the proposed AF can solve the issue of many active sources, we increase the the number of active sources to $N=10$ with the incident angles spread from $-60^o$ to $60^o$. 
As the results are shown in Fig.~\ref{fig:f2}, both the MUSIC baseline and proposed AF work well, record the RMSE scores of $0.23^o$ and $0.5^o$ , respectively.
\begin{figure}[t]
	\begin{center}
		\includegraphics[width=9cm]{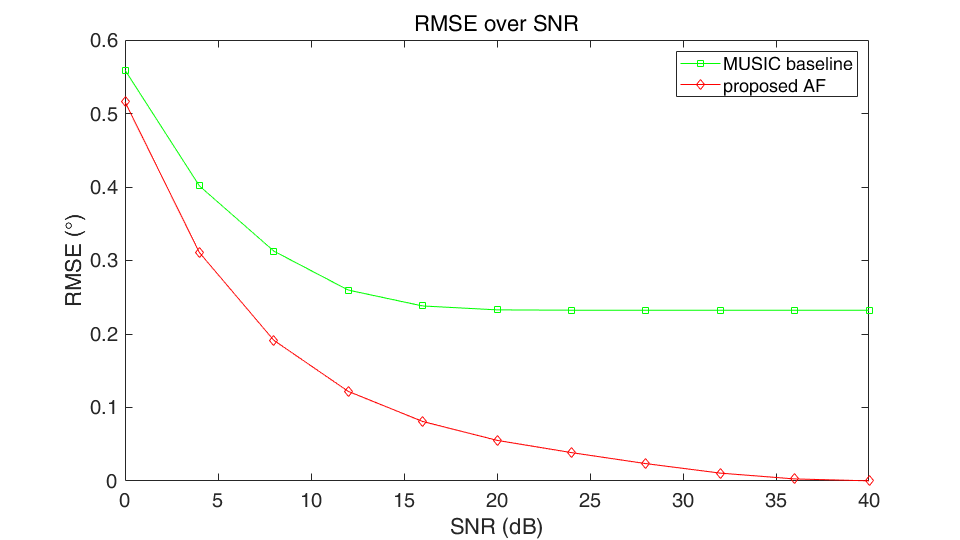}
	\end{center}
    \vspace{-0.4 cm}
	\caption{Performance comparison between MUSIC baseline and proposed AF-based method with $N =10$, white noise, 1000 Monte Carlo trials.}
	     \vspace{-0.4 cm}
	\label{fig:f3}
\end{figure}
\begin{figure}[t]
	\begin{center}
		\includegraphics[width=\linewidth]{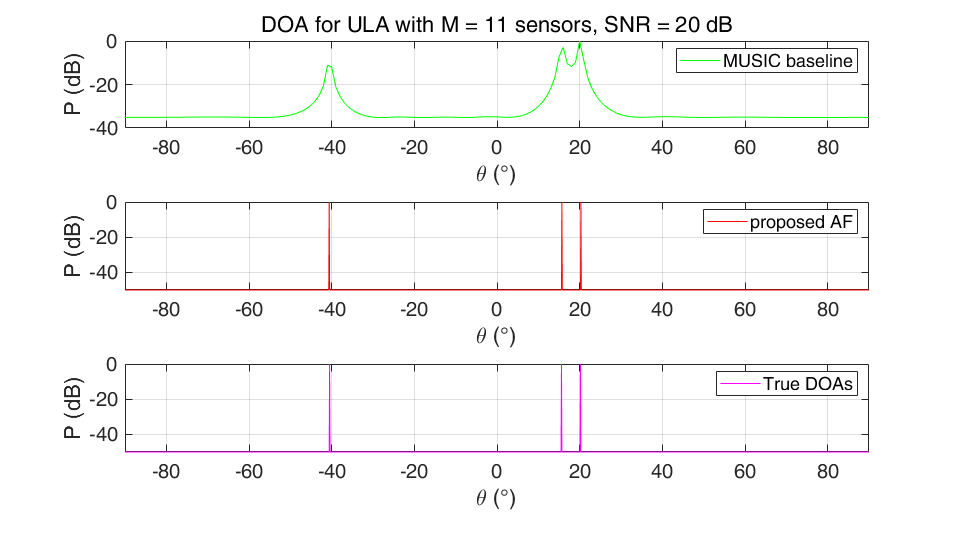}
	\end{center}
    \vspace{-0.6 cm}
	\caption{Power spectrum comparison with SNR = 20 dB, white noise, N = 3.}
	     \vspace{-0.4 cm}
	\label{fig:f4}
\end{figure}

To compare the performance between the MUSIC baseline and proposed AF, we conduct an experiment of 1000 Monte Carlo trials with the same setting of N = 10 and white noise only.
As shown in Fig.~\ref{fig:f3}, it can be seen that the proposed AF method outperforms the MUSIC baseline in wide range of SNR.
To further evaluate the MUSIC baseline and proposed AF, we conducted experiment with the setting: SNR = 20 dB, the number of active sources $N=3$ with the incident angles of $-40.5^o$, $15.6^o$ and $20.2^o$ respectively.
As the results of spectrum power shows in Fig.~\ref{fig:f4},  while the MUSIC-based baseline detects the arrived signal from $-40^o$, $16^o$ and $20^o$, the proposed AF detects three sources at  $-40.5378^o$,  $15.6486^o$ and $20.2451^o$.
It can be seen that the proposed AF achieves the higher accuracy, improves the the MUSIC-based baseline $0.5^o$ in term of RMSE score.
The lower performance of the MUSIC baseline can be explained by searching grid of MUSIC algorithm, leading the dependence of grid resolution (e.g. the grid resolution is set to $1^o$ in our experiments).

\subsection{Simulations With Both White Noise and Diffuse Noise}
\begin{figure}[t]
	\begin{center}
		\includegraphics[width=9cm]{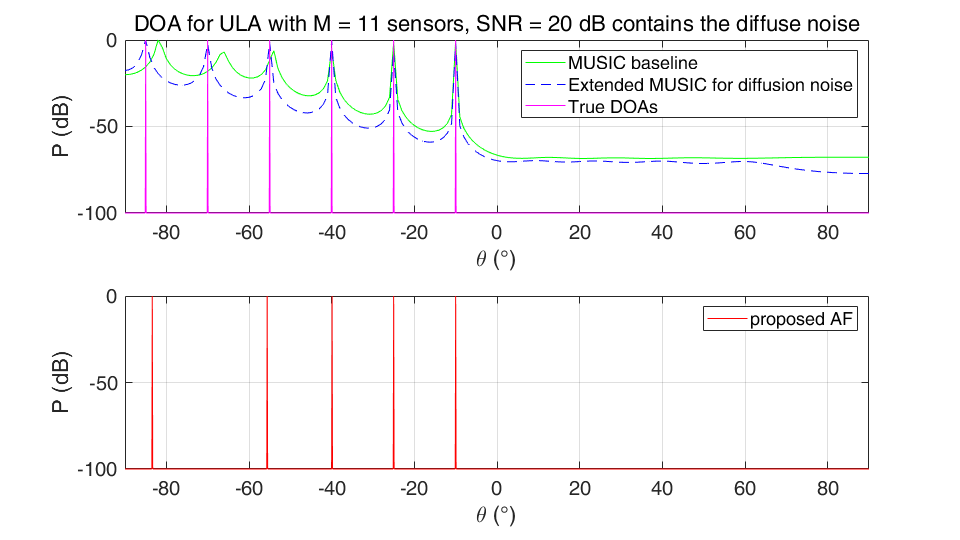}
	\end{center}
     \vspace{-0.6 cm}
	\caption{Power Spectrum of different methods (SNR = 20 dB contains the white noise and diffuse noise): MUSIC, extended MUSIC for diffuse noise, and the true DOAs are presented in the top figure, the blow is for the AFM.}
	\label{fig:DOA_compare_spectra_4}
\end{figure}

Considering both white noise and diffusion noise with ${\sigma^2_d / \sigma^2_w} = 25$. The other settings are SNR = 20 dB, the number of sources $N = 5$.
Also, the inter-distance of sensors is reduced to less than half of the wavelength to achieve a reasonable diffuse noise correlation matrix (the off-diagonal elements of $\mathbf{\Gamma}$ are not $0$).
We compare the proposed AF with the MUSIC baseline and extended MUSIC for diffusion noise. Only the extended MUSIC for diffusion noise can estimate the DOAs properly, as shown in Fig.~\ref{fig:DOA_compare_spectra_4}.  The RMSEs of MUSIC baseline, extended MUSIC for diffuse noise and proposed AF are $2.1^o$, $0.0^o$ and $28.6^o$, respectively.  

\section{Conclusions}
\label{sec:C6_Conclusion}
In this paper, we have proposed an annihilating filter-based technique for DOA estimation. The proposed method processes on multiple frames under the constrain of frame-variant or incoherent signals. The maximum number of detectable sources is almost twice times of that of conventional annihilating filter-based DOA estimation. In comparison with MUSIC, the proposed method is independent with the grid directions, then its performance outperforms the MUSIC algorithm in terms of accuracy. Moreover, the complexity of new method is $\mathcal{O}(KM^2)$, which is less than the complexity of subspace-based techniques. However, when the diffuse noise presents in the measurement signal, only extended MUSIC, which is also newly proposed in this paper, could estimate the DOA properly.


\bibliographystyle{IEEEtran}
\bibliography{chapter6}


\end{document}